\documentclass[conference]{IEEEtran}
\usepackage{cite}
\usepackage{amsmath,amssymb,amsfonts}
\usepackage{graphicx}
\usepackage{textcomp}
\usepackage{xcolor}
\def\BibTeX{{\rm B\kern-.05em{\sc i\kern-.025em b}\kern-.08em
    T\kern-.1667em\lower.7ex\hbox{E}\kern-.125emX}}

\usepackage{algpseudocode}
\usepackage{booktabs} 

\usepackage{tikz}
\usetikzlibrary{patterns}
\usetikzlibrary{mindmap}
\usetikzlibrary{calc}

\usepackage{pgfplots}
\pgfplotsset{compat=1.13}

\usepackage{adjustbox}

\usepackage{hyperref}

\usepackage{listings}
\newtheorem{theorem}{Theorem}
\newtheorem{lemma}{Lemma}

\DeclareMathOperator*{\argmin}{arg\,min}
\DeclareMathOperator*{\argmax}{arg\,max}

\newcommand{\BigO}{\mathop{\text{O}}}

\newcommand{\Th}{\mathop{\Theta}}

\newcommand{\GF}{\operatorname{GF}}
\newcommand{\F}{\mathbb{F}}
\newcommand{\E}{\operatorname{E}}

\tikzset{%
    vertex/.style={circle,draw=blue!50,fill=blue!20,thick,minimum size=6mm},
    vertex+/.style={circle,draw=blue!50,fill=blue!20,thick,minimum size=8mm,text=orange},
    bubble/.style={circle,inner sep=0,minimum size=7.5mm,fill=orange,draw=orange,thick},
    concept color=orange,
    decoration={start radius=3.5mm,end radius=3.5mm,amplitude=3mm,angle=40},
    bubbleedge/.style={draw=orange,ultra thin,fill=orange,decorate,decoration=circle connection bar},
    dot/.style={circle,fill=blue!50,inner sep=0,minimum size=2mm},
    edge/.style={thick},
    arc/.style={thick,->},
    rep/.style={orange},
    newedge/.style={green!70!black},
	rep2/.style={red}
}

\begin{document}

\title{In-database connected component analysis}

\author{\IEEEauthorblockN{Harald B\"ogeholz}
\IEEEauthorblockA{\textit{Faculty of Information Technology} \\
\textit{Monash University}\\
Melbourne, Australia \\
harald.boegeholz@monash.edu}
\and
\IEEEauthorblockN{Michael Brand}
\IEEEauthorblockA{\textit{Otzma Analytics Pty Ltd} \\
Melbourne, Australia \\
research@OtzmaAnalytics.com}
\IEEEauthorblockA{\textit{Faculty of Information Technology} \\
\textit{Monash University}\\
Melbourne, Australia}
\and
\IEEEauthorblockN{Radu-Alexandru Todor}
\IEEEauthorblockA{\textit{UBS}\\
Z\"urich, Switzerland \\
radu-alexandru.todor@ubs.com}
}

\maketitle

\begin{abstract}
We describe a Big Data-practical, SQL-im\-ple\-mentable algorithm for efficiently
determining connected components for graph data stored in a Massively Parallel Processing (MPP) relational database.
The algorithm described is a linear-space, randomised algorithm, always
terminating with
the correct answer but subject to a stochastic running time, such that for any
$\epsilon>0$ and any input graph $G=\langle V, E \rangle$ the algorithm
terminates after $\mathop{\text{O}}(\log |V|)$ SQL queries with probability of
at least $1-\epsilon$, which we show empirically to translate to a quasi-linear runtime in practice.\end{abstract}


\section{Introduction}

Connected component analysis \cite{Hopcroft:CC}, the assignment of a label to each vertex in a graph such that two vertices receive the same label if and only if they belong to the same connected component, is one of the tent-pole algorithms of graph analysis. Its wide use is in applications ranging from image processing (e.g., \cite{kikuchi:spots,song:region,nowosielski:vehicle,wu:nests}, to name a few recent examples) to cyber-security (e.g., \cite{Giani:grids,Patil:governance,Hogan:multiscale,Yip:criminal}). The most well-known theoretical result regarding connectivity analysis is perhaps the Union/Find algorithm \cite{Hopcroft:merging,Tarjan:efficiency,Cormen:disjoint}, ensuring that labels can be maintained per vertex in an amortised complexity on the order of the inverse Ackermann function per edge, which is the theoretical optimum. 

In real-world settings, however, large graphs such as those analysed in Big Data data science are stored on distributed file systems and processed in distributed computing environments. These are ill-suited for the Union/Find algorithm. For example, Union/Find involves following long linked lists, which is inefficient if the items in these lists reside on different machines.

A widely used platform for Big Data processing is Hadoop with its distributed and redundant file system HDFS and the MapReduce framework for implementing distributed computation~\cite{miner2012mapreduce}. Another, more recent distributed computing framework is Apache Spark~\cite{karau2017high}, building on Hadoop HDFS for data storage. These two have in common that algorithms have to be specifically designed for the respective framework.
 
However, most of the world's transactional business data is stored natively in large, relational, SQL-accessible databases, and is only treated as graph data in certain contexts. It is therefore beneficial to have an efficient solution for graph algorithms, and particularly for the connected components algorithm, within the framework of relational databases. Such a solution obviates the need for data duplication in a separate storage system and for supporting multiple data storage architectures. It also avoids the potential for data conflicts and other problems arising from performing data analysis in two disparate systems.

The present paper presents a new algorithm for connected components analysis, Randomised Contraction. It is practical for Big Data data analytics in the following respects:

\begin{LaTeXdescription}

\item[In-database execution.] Our algorithm uses SQL queries as its basic building blocks. It can therefore be natively executed in a relational database, and specifically within the framework of Massively Parallel Processing (MPP) databases \cite{Dewitt:MPP} where the architecture is designed for efficient parallel processing. 

\item[Scalability.] Randomised Contraction uses (for any input graph) an expected logarithmic number of queries, running over exponentially decreasing amounts of data. Our empirical results obtained with an MPP database show it to smoothly scale out to Big Data, running, in total, in an amount of time quasi-linear in the input size.

\item[Space efficiency.] Typical database maintenance uses some bounded fraction of available space. Therefore, practical in-database algorithms for use on mass data should not create intermediate data that is more than linear in the size of the input. Our algorithm satisfies this criterion.

\end{LaTeXdescription}

Our empirical results show that Randomised Contraction outperforms other leading connected components algorithms when implemented in an MPP database. Furthermore, our in-database implementation of one of the algorithms runs faster than the original Spark implementation and uses fewer resources, allowing it to scale up to larger datasets.

The paper is structured as follows: Section~\ref{S:related} presents related work. In Section~\ref{S:problem}, we describe the problem formally. In Section~\ref{S:simple}, we discuss naive approaches to a solution and show where they fail. In Section~\ref{S:new}, we describe our new algorithm, Randomised Contraction, with several refinements, and in Section~\ref{S:performance} we analyse its theoretical performance. Section~\ref{S:empirical} gives empirical results. A short conclusions section follows. Appendix~\ref{A:implementation} presents excerpts of the code used for experiments. Appendix~\ref{A:bounds} gives improved theoretical bounds on graph contraction that may be of independent interest.

\section{Related work}\label{S:related}

Many researchers have long tried to optimise connected component finding for parallel computing environments (e.g., \cite{Hirschberg:parallel}). Most suited for this pursuit from a theoretical perspective is the theoretical framework of the Parallel Random Access Machine (PRAM) \cite{Fich:PRAM,Savage:models}. PRAM algorithms for connected components finding were presented, e.g., in \cite{Shiloach:Ologn,Vishkin:optimal,Awerbuch:1987}. In \cite{Gazit:randomized}, it was noted that randomised algorithms may have an advantage in this problem. The best result obtained by the randomised approach is \cite{Halperin:optimal}, where a randomised EREW PRAM algorithm is presented that finds connected components of a graph $G=\langle V,E \rangle$ in $\BigO(\log |V|)$ time using an optimal number of $\BigO((|V|+|E|)/\log |V|)$ processors. Its result is always correct and the probability that it does not complete in $\BigO(\log |V|)$ time is at most $n^{-c}$ for any $c>0$.

However, as observed by Eppstein and Galil \cite{Eppstein:PRAMs}, the PRAM model is ``commonly used by theoretical computer scientists but less often by builders of actual parallel machines''. Its assumptions, which are idealisations of the parallelised computation set-up, do not accurately reflect the realities of parallel computing architectures, making its algorithms unrealistic to implement or not truly attaining the reported performance complexity bounds.

Indeed, the papers that explore connected components algorithms for large-scale practical architectures do so using decidedly different algorithms. The first MapReduce algorithms that run in a logarithmic number of rounds were proposed by Rastogi et al.~\cite{Rastogi:MRCC}. Among several variations of new algorithms presented, they report the overall best practical performance for the Hash-to-Min algorithm. This algorithm, however, has a worst case space usage of $\BigO(|V|^2)$. The best known space usage of a MapReduce algorithm is linear in the input size and achieved by the Tho-Phase algorithm by Kiveris et al.~\cite{kiveris2014connected}. This algorithm, however, takes $\Th(\log^2|V|)$ rounds. The Cracker algorithm proposed by Lulli et al.~\cite{lulli2017fast} is implemented in Spark and once again improves the number of rounds to $\BigO(|V|)$, but it does so at the expense of increasing the communication cost to $\BigO(\frac{|V|\cdot|E|}{\log|V|})$. 

As outlined in the introduction, if the data to be analysed is already stored in a distributed relational database, it is beneficial to be able to run algorithms in-database instead of exporting data to a different platform for analysis. This led to the development of the open source machine learning library Apache MADlib~\cite{hellerstein2012madlib}. This library implements, among a small set of other graph algorithms, a connected components algorithm using Breadth First Search. We show in section~\ref{S:simple} that its worst case behaviour makes it unsuitable for Big Data data science.

Our novel Randomised Contraction algorithm has an efficient implementation in an MPP database and achieves both the best time complexity and space complexity among the above mentioned algorithms. Like the PRAM algorithms of \cite{Gazit:randomized,Halperin:optimal}, it is randomised. It is guaranteed to terminate and to do so with a correct answer, and for any given $\epsilon>0$ guarantees to terminate after $\BigO(\log |V|)$ SQL queries with probability at least $1-\epsilon$, where $|V|$ is the number of vertices in the input graph. The algorithm's space requirements can be made linear deterministically, not merely in expectation, and it can be implemented to use temporary storage not exceeding four times the size of the input plus $\BigO(|V|)$. This is at worst a five-fold blow-up, which is within the typical range for standard database operations.

\section{Problem description}\label{S:problem}

A graph $G=\langle V,E \rangle$ is typically stored in a relational database in the form of two tables. One stores the set of vertices $V$, represented by a column of unique vertex IDs and optionally more columns with additional vertex information. Another table stores the edge set $E$ in two columns containing vertex IDs and optionally more columns with additional edge information. In the context of connected component analysis, graphs are taken to be undirected, so an $(x,y)$ edge is considered identical to a $(y,x)$ edge. For simplicity we present our algorithm such that its only input is an edge table containing two columns with vertex IDs from which the set of vertices is deduced. Isolated vertices can be represented in this table as ``loop edges'', $(v,v)$, if necessary.

The output of the algorithm is a single table with two columns, $v$ and $r$, containing one row per vertex. In each row $v$ is a vertex ID and $r$ is a label uniquely identifying the connected component the vertex belongs to. A correct output of the algorithm is one where any two vertices share the same $r$ value if and only if they belong to the same connected component. Connected component analysis does not make any specific requirement regarding the values used to represent components other than that they are comparable.

\section{Simple solution attempts}\label{S:simple}

Perhaps the simplest approach to performing in-database connected components
analysis is to begin by choosing for each vertex a representative by picking the vertex with the minimum ID among the vertex itself and all its neighbours, then to improve on that representative by taking
the minimum ID among the \emph{representatives} of the vertex itself and
all its neighbours, and to continue in this fashion until no vertex
changes its choice of representative. We refer to this naive approach as the ``Breadth First Search'' strategy:
after $n$ steps each vertex's representative is the vertex with the
minimum ID among all vertices in the connected component that are at most
at distance $n$ from the original vertex.

Though the algorithm ultimately terminates and delivers the correct
result, its worst-case runtime makes it unsuitable
for Big Data. Consider, for example, the sequentially numbered path graph with IDs $1$, $2$, \dots, $n$. For this graph, Breadth First Search will take $n-1$ steps.

To remedy this, consider an
algorithm that calculates $G^2$, i.e.\ the graph over the same vertex set
as $G$ whose set of edges includes, in addition to the original edges, also $(x,z)$ for every $x$ and $z$ for which
there exists a $y$ such that both $(x,y)$ and $(y,z)$ are edges in $G$.

Calculating $G^2$ can be done easily in SQL by means of a self-join. A tempting
possibility is therefore to repeat the self-join and calculate
$G^4$, $G^8$, etc.. Such a procedure would allow us to reach neighbourhoods
of radius $2^n$ in only $n$ steps.

Unfortunately, this second approach does not yield a workable algorithm,
either. The reason for this is that in $G^k$ each vertex is directly
connected to its entire neighbourhood of radius $k$ in~$G$. For a single-component $G$, the result is ultimately the complete graph with $|V|^2$ edges. This is a quadratic blow-up in data size, which for Big Data analytics is unfeasible.

Our aim, in presenting a new algorithm, is therefore to enjoy the best of both worlds: we would like to be able to work in a number of operations logarithmic in the size of the graph, but to require only linear-size storage.

\section{The new algorithm}\label{S:new}

We present our new algorithm for calculating connected components, \textbf{Randomised Contraction}, by starting with its basic idea and refining it in several steps.

\subsection{The basic idea}
\label{S:basic}

\begin{figure*}
\footnotesize
\setlength{\tabcolsep}{3.5pt}
\begin{tabular}{cccccc}
\begin{adjustbox}{valign=t}
\begin{tikzpicture}[scale=0.7]

  \node (3) at (0,0) [vertex] {3};
  \node (1) at (-0.5,2.5) [vertex] {1};
  \node (6) at (1.5,2.5) [vertex] {6};
  \node (7) at (1.5,4.5) [vertex] {7};
  \node (8) at (1.5,0) [vertex] {8};
  \node (10) at (0.5,1.5) [vertex] {10};
  \node (5) at (0.5,3.5) [vertex] {5};
  \node (2) at (4,3) [vertex] {2};
  \node (9) at (3,1.5) [vertex] {9};
  \node (4) at (5,1.5) [vertex] {4};

  \begin{scope}[edge]
      \draw (1) -- (5);
      \draw (1) -- (10);
      \draw (2) -- (4);
      \draw (2) -- (9);
      \draw (3) -- (8);
      \draw (3) -- (10);
      \draw (4) -- (9);
      \draw (5) -- (7);
      \draw (6) -- (5);
      \draw (6) -- (10);  
  \end{scope}
  
\end{tikzpicture}
\end{adjustbox}
&
\begin{tabular}[t]{cc}
$v$ & $w$ \\
\midrule
1 & 5 \\
1 & 10 \\
2 & 4 \\
2 & 9 \\
3 & 8 \\
3 & 10 \\
4 & 9 \\
5 & 6 \\
5 & 7 \\
6 & 10 \\
\end{tabular}
&
    \begin{tabular}[t]{cc}
    $x$ & $r_1(x)$ \\
	\midrule
    1 & 1 \\
    2 & 2 \\
    3 & 3 \\
    4 & 2 \\
    5 & 1 \\
    6 & 5 \\
    7 & 5 \\
    8 & 3 \\
    9 & 2 \\
    10 & 1
    \end{tabular}
&
\pgfdeclarelayer{bubbles}
\pgfsetlayers{bubbles,main}
\begin{adjustbox}{valign=t}
\begin{tikzpicture}[scale=0.7]

  \node (3) at (0,0) [vertex] {3};
  \node (1) at (-0.5,2.5) [vertex] {1};
  \node (6) at (1.5,2.5) [vertex] {6};
  \node (7) at (1.5,4.5) [vertex] {7};
  \node (8) at (1.5,0) [vertex] {8};
  \node (10) at (0.5,1.5) [vertex] {10};
  \node (5) at (0.5,3.5) [vertex] {5};
  \node (2) at (4,3) [vertex] {2};
  \node (9) at (3,1.5) [vertex] {9};
  \node (4) at (5,1.5) [vertex] {4};
  \begin{scope}[edge]
      \draw (1) -- (5);
      \draw (1) -- (10);
      \draw (2) -- (4);
      \draw (2) -- (9);
      \draw (3) -- (8);
      \draw (3) -- (10);
      \draw (4) -- (9);
      \draw (5) -- (7);
      \draw (6) -- (5);
      \draw (6) -- (10);
  \end{scope}

  \begin{pgfonlayer}{bubbles}
    	\foreach \i in {1,...,10} {
    		\node (b\i) at (\i) [bubble] {};
    	}
    	\filldraw [bubbleedge] (b1)--(b10);
    	\filldraw [bubbleedge] (b1)--(b5);
    	\filldraw [bubbleedge] (b6)--(b7);
    	\filldraw [bubbleedge] (b3)--(b8);
    	\filldraw [bubbleedge] (b2)--(b4);
    	\filldraw [bubbleedge] (b4)--(b9);
    	\filldraw [bubbleedge] (b9)--(b2);

		\node [bubble,minimum size=9mm] at 
			(barycentric cs:b2=1,b4=1,b9=1) {}; 
  \end{pgfonlayer}

    \node[rep,right] at (b7.east) {5};
    \node[rep,left] at (b5.west) {1};
    \node[rep,left] at (b1.west) {1};
    \node[rep,right] at (b6.east) {5};
    \node[rep,left] at (b10.west) {1};
    \node[rep,above right] at (3.north east) {3};
    \node[rep,above right] at (8.north east) {3};
    
    \node[rep,right] at (b2.east) {2};
    \node[rep,below right] at (9.south east) {2};
    \node[rep,below right] at (4.south east) {2};

\end{tikzpicture}
\end{adjustbox}
&
	\begin{tabular}[t]{cccc}
&    $r_1(v)$ & $r_1(w)$ &\\
    \midrule
&	1 & 1 &\\
    \noalign{\vspace{-1.5ex}}\cline{2-3}\noalign{\vspace{\dimexpr 1.5ex}}
&	1 & 1 &\\
    \noalign{\vspace{-1.5ex}}\cline{2-3}\noalign{\vspace{\dimexpr 1.5ex}}
&	2 & 2 &\\
    \noalign{\vspace{-1.5ex}}\cline{2-3}\noalign{\vspace{\dimexpr 1.5ex}}
&	2 & 2 &\\
    \noalign{\vspace{-1.5ex}}\cline{2-3}\noalign{\vspace{\dimexpr 1.5ex}}
&	3 & 3 &\\
    \noalign{\vspace{-1.5ex}}\cline{2-3}\noalign{\vspace{\dimexpr 1.5ex}}
&	3 & 1 &\\
&	2 & 2 &\\
    \noalign{\vspace{-1.5ex}}\cline{2-3}\noalign{\vspace{\dimexpr 1.5ex}}
&	1 & 5 &\\
&	5 & 5 &\\
    \noalign{\vspace{-1.5ex}}\cline{2-3}\noalign{\vspace{\dimexpr 1.5ex}}
&	5 & 1 &\\
    \noalign{\vspace{-1.5ex}}\cline{2-3}\noalign{\vspace{\dimexpr 1.5ex}}
	\end{tabular}
&
\begin{adjustbox}{valign=t}
\begin{tikzpicture}[scale=0.7]

    \node (5) at (0.5,4) [vertex] {5};
    \node (1) at (0,2) [vertex] {1};
    \node (3) at (0.5,0) [vertex] {3};
	\node (2) at (2,2) [vertex] {2};
	
    \draw[edge] (1) -- (5);
    \draw[edge] (1) -- (3);
	\draw ($(2.west)+(-6pt,0)$) -- ($(2.east)+(6pt,0)$);

\end{tikzpicture}
\end{adjustbox}
\\
\noalign{\vskip5pt}
(a)&(b)&(c)&(d)&(e)&(f)\\
\end{tabular}
\caption{(a) An undirected graph $G_0$ with vertex IDs shown inside the nodes. (b) The representation of $G_0$ as a list of edges. (c) The choice of representative $r_1(x)$ for each vertex $x$. (d) The graph with representative choices shown at the side of each node. Bubbles around the nodes indicate sets of vertices with the same choice of representative. These will be contracted to single vertices. (e) The edge list of the graph $G_1$ is computed by mapping the function $r_1$ over the edge list of $G_0$. Duplicates and loop edges, shown struck out, are eliminated. (f) The resulting graph $G_1$ after one contraction step. The isolated vertex~2, shown struck out, is excluded from further computation.}
\label{F:contraction}
\end{figure*}
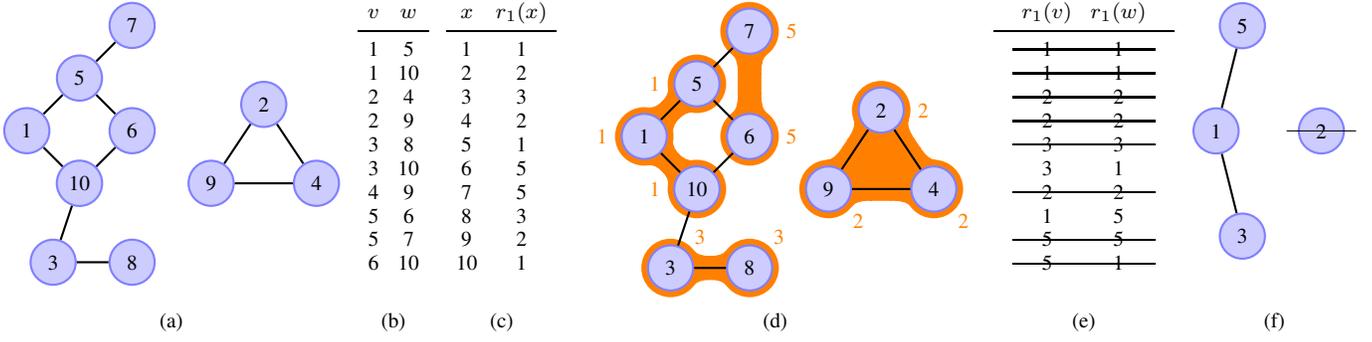

\begin{figure}
\footnotesize
\pgfdeclarelayer{bubbles}
\pgfsetlayers{bubbles,main}
\begin{tabular}{ll}
(a) &
    \begin{tikzpicture}[scale=0.7,baseline=(1.base)]
    \node (1) [vertex] at (1.5,0) {1};
    \node (2) [vertex] at (3.0,0) {2};
    \node (3) [vertex] at (4.5,0) {3};
    \node (4) [vertex] at (6.0,0) {4};
    \node (5) [vertex] at (7.5,0) {5};
    \node (6) [vertex] at (9.0,0) {6};
    
    \draw [edge] (1) -- (2) -- (3) -- (4) -- (5) -- (6);
    
    \begin{pgfonlayer}{bubbles}
    	\foreach \i in {1,...,6} {
    		\node (b\i) at (\i) [bubble] {};
    	}
    	\filldraw [bubbleedge] (b1)--(b2);
	\end{pgfonlayer}

    \node[rep,above] at (b1.north) {1};
    \node[rep,above] at (b2.north) {1};
    \node[rep,above] at (b3.north) {2};
    \node[rep,above] at (b4.north) {3};
    \node[rep,above] at (b5.north) {4};
    \node[rep,above] at (b6.north) {5};

    \end{tikzpicture}
\\[12pt]
(b) &
    \begin{tikzpicture}[scale=0.7,baseline=(1.base)]
    \node (1) [vertex] at (1.5,0) {3};
    \node (2) [vertex] at (3.0,0) {1};
    \node (3) [vertex] at (4.5,0) {4};
    \node (4) [vertex] at (6.0,0) {5};
    \node (5) [vertex] at (7.5,0) {2};
    \node (6) [vertex] at (9.0,0) {6};
    
    \draw [edge] (1) -- (2) -- (3) -- (4) -- (5) -- (6);
    
    \begin{pgfonlayer}{bubbles}
    	\foreach \i in {1,...,6} {
    		\node (b\i) at (\i) [bubble] {};
    	}
    	\filldraw [bubbleedge] (b1)--(b2);
    	\filldraw [bubbleedge] (b2)--(b3);
    	\filldraw [bubbleedge] (b4)--(b5);
    	\filldraw [bubbleedge] (b5)--(b6);
	\end{pgfonlayer}

    \node[rep,above] at (b1.north) {1};
    \node[rep,above] at (b2.north) {1};
    \node[rep,above] at (b3.north) {1};
    \node[rep,above] at (b4.north) {2};
    \node[rep,above] at (b5.north) {2};
    \node[rep,above] at (b6.north) {2};

    \end{tikzpicture}
\\
\end{tabular}
\caption{(a) In a sequentially numbered path graph, every vertex but the first one will choose its left neighbour as a representative. This is the worst case: the contracted graph is only one vertex smaller. (b) If the same path graph is numbered optimally, it contracts to $1/3$ the number of vertices.}
\label{F:pathcontract}
\end{figure}
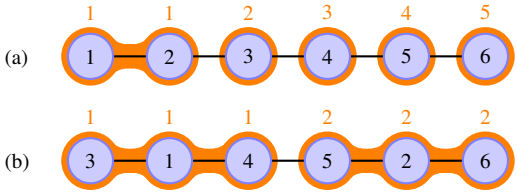

Let $G=\langle V, E\rangle$ be a graph. The algorithm contracts the graph to a set of representative vertices, preserving connectivity, and repeats that process until only isolated vertices remain. These then represent the connected components of the original graph.

Denote by $N_G[v]$ the \emph{closed neighbourhood} of a vertex $v$, i.e.\ the set of all vertices connected to $v$ by an edge in $E$ plus $v$ itself. Let $G_0=\langle V_0, E_0 \rangle$ be the original graph. 

At step $i$, map every vertex $v$ to a \emph{representative} $r_i(v) \in N_{G_{i-1}}[v]$. The contracted graph $G_i = \langle V_i, E_i \rangle$ is then constructed as $V_i=\{r_i(v)\mid v\in V_{i-1}\}$ and $E_i=\{(r_i(v),r_i(w))\mid (v,w)\in E_{i-1} \text{ and } r_i(v)\ne r_i(w)\}$. Note that two vertices are connected in $G_{i-1}$ if and only if their representatives are connected in $G_i$. In other words, for each connected component of $G_{i-1}$ there is a corresponding connected component in $G_i$.

Repeat this contraction process until reaching a graph $G_k$ that contains only isolated vertices. At that stage each of these represents one of the connected components of the original graph. Applying all the maps $r_i$ in sequence maps each vertex to an identifier unique to its connected component: the composition of the representative functions $r_k \circ r_{k-1} \circ \cdots \circ r_1$ is the output of the algorithm.

Assuming the vertices are ordered, the basic idea for the choice of representatives is to set $r_i(v) = \min N_{G_{i-1}}[v]$. After each contraction step, isolated vertices can be excluded from further computation since each of them is known to form a connected component by itself. If the graph is only represented by its edge set, the removal of loop edges effectively eliminates isolated vertices. This leads to a natural termination condition: the algorithm terminates when the edge set becomes empty. Figure~\ref{F:contraction} illustrates one contraction step using this idea. The graph (a) is represented as a list of edges (b). The edge list of the contracted graph (e) is obtained by mapping the representative function over all vertex IDs in this list, eliminating duplicates and eliminating loop edges.

\subsection{Randomisation}
\label{S:randomisation}

The algorithm in the previous section still suffers from the same worst case as the Breadth First Search strategy described in Section~\ref{S:simple}. Consider a sequentially numbered path graph on $n$ vertices as shown in Fig.~\ref{F:pathcontract}(a). Each vertex except the first one will choose as its representative the neighbour preceding it. The result of contraction is a sequentially numbered path graph on $n-1$ vertices. This implies that the algorithm takes $n-1$ steps until the path is contracted to a single vertex. If, on the other hand, the path is labelled differently, it can contract to $1/3$ of its vertices in the optimal case as shown in Fig.~\ref{F:pathcontract}(b).

A solution for avoiding worst case contraction is to randomise the order of the vertices. We show in Section~\ref{S:performance} that the graph will then, in expectation, shrink to at most a constant fraction $\gamma$ of its vertices, with $\gamma<1$. We further show that if the randomisation is performed independently at each step, this leads to an expected logarithmic number of steps. As a result, the algorithm behaves well for any input. By contrast, other algorithms that rely on a worst case being ``unlikely'' are vulnerable in an adversarial scenario where such a worst case can be exploited to an attacker's advantage.

We remark that vertex label randomisation, critical to our algorithm, would not have aided the simple solution attempts described in Section~\ref{S:simple}. The complexity of Breadth First Search, for example, is bounded by the diameter of the analysed graph, regardless of how vertices are labelled. 

\subsection{Randomisation methods}

In a practical implementation, choosing a random permutation of the vertices is itself a nontrivial task, especially in a distributed computing scenario such as an MPP database. One way to achieve this is the \textbf{random reals method}. At step $i$, generate for each vertex $v$ a random real $h_i(v)$ uniformly distributed in $[0,1]$. The choice of the representative then becomes $r_i(v) = \argmin_{w\in N_{G_{i-1}}[v]} h_i(w)$.

This method in theory achieves \emph{full randomisation}, a uniform choice among all $|V|!$ possible orderings of the vertices, for which the best performance bounds can be proved (see Appendix~\ref{A:bounds}). The advantage of the random reals method over brute-force random permutation generation is that the table of random numbers can be created in parallel in a distributed database. A disadvantage is that this table has to be distributed to all machines in the cluster for picking representatives.

A more efficient idea is to pick a pseudo-random permutation by means of an encryption function on the domain of the vertex IDs. If the vertex IDs are 64-bit integers, a suitable choice is the Blowfish algorithm~\cite{schneier1993description} which can be implemented in a database as a user-defined function. Let $e_k$ denote an encryption function on the domain of the vertex IDs with key $k$. The \textbf{encryption method} then works as follows: at step $i$, choose a random key $k_i$. Let $r_i(v) = \argmin_{w\in N_{G_{i-1}}[v]} e_{k_i}(w)$. Note that an encryption function is by definition a bijection which guarantees a unique choice of representatives.

The encryption method is more efficient than the random reals method in a distributed setting since it obviates the need to communicate one random number per vertex across the network to every node that needs it. Instead, only the encryption key needs to be distributed and each processor can compute the pseudo-random vertex IDs independently as necessary. This exploits the fact that in a realistic setting, communication across computation nodes is much slower than local computing.

While encryption functions are designed to be ``as random as possible'' and work well in practice, it is hard to rigorously prove for them the required graph contraction properties. Also, they are computationally expensive. We therefore present as the final refinement of the Randomised Contraction algorithm the \textbf{finite fields method}.  Assume the domain of the vertex IDs is a finite field $\F$ with any ordering. To determine the representatives at step $i$, choose $0\ne A_i\in \F$ and $B_i\in \F$ uniformly at random and let $r_i(v) = \argmin_{w\in N_{G_{i-1}}[v]} h_i(w)$ where $h_i(w) = A_i \cdot w + B_i$ with multiplication and addition carried out using finite field arithmetic. Note that $h_i$ is a bijection: in a field, every $A \ne 0$ has a unique multiplicative inverse $A^{-1}$. If $y=A\cdot x+B$, we have $x=A^{-1}\cdot(y-B)$.

If the vertex IDs are fixed-size integers with $b$ bits, this data type can be treated as a finite field with $2^b$ elements by performing polynomial arithmetic modulo an irreducible polynomial~\cite[Thm.~3.2.6]{ling2004coding}. Note that while the calculation of $h_i(w)$ is performed in the finite field $\F$, the result is stored as an integer and the calculation of $\argmin$ is done with reference to integer ordering. Since finite field arithmetic over this field is awkward to implement in SQL, we wrote a fast implementation in C and loaded it as a user-defined function into the database. An SQL-only implementation could alternatively choose a prime number $p$ known to be larger than any vertex ID and use normal integer arithmetic modulo $p$, giving the data type of the vertex IDs the structure of $\F=\GF(p)$.

\subsection{SQL implementation}\label{S:implementation}

\begin{figure}
\begin{algorithmic}
\footnotesize 
\Procedure{RandomisedContraction}{$G$}
	\State \textbf{create table E as}
	\State \hskip\algorithmicindent \textbf{select v, w from $G$ union all select w, v from $G$;}
	\State $\textit{firstround} \gets \textit{true}$
	\Repeat
		\State choose $0 \ne A \in \F$ and $B \in \F$ uniformly at random
		\State \textbf{create table R as}\Comment{compute representatives}
		\State \hskip\algorithmicindent \textbf{select v, least(axb($A$, v, $B$), min(axb($A$, w, $B$))) as r}
		\State \hskip\algorithmicindent \textbf{from E group by v;}
		\State \textbf{create table T as}\Comment{contract by transforming edge table}
		\State \hskip\algorithmicindent \textbf{select distinct V.r as v, W.r as w}
		\State \hskip\algorithmicindent \textbf{from E, R as V, R as W}
		\State \hskip\algorithmicindent \textbf{where E.v = V.v and E.w = W.v and V.r != W.r;}
		\State \textit{rowcount} $\gets$ number of rows generated by the previous query
		\State \textbf{drop table E; alter table T rename to E;}
		\If{\textit{firstround}}
			\State $\textit{firstround} \gets \textit{false}$
			\State \textbf{alter table R rename to L;}
		\Else
    		\State \textbf{create table T as}\Comment{compose representative functions}
    		\State \hskip\algorithmicindent \textbf{select L.v as v, coalesce(R.r, axb($A$, L.r, $B$)) as r}
    		\State \hskip\algorithmicindent \textbf{from L left outer join R on (L.r = R.v);}
    		\State \textbf{drop table L, R; alter table T rename to L;}
		\EndIf
	\Until{$\textit{rowcount} = 0$}
	\State \textbf{alter table L rename to Result;}
\EndProcedure
\end{algorithmic}
\caption{SQL-like pseudocode for the Randomised Contraction algorithm with deterministic space usage using the finite fields method. axb is assumed to be a user-defined function that computes the term $A\cdot x+B$ using arithmetic over the finite field~$\F$.}
\label{F:sqlrc}
\end{figure}

Our implementation of the Randomised Contraction algorithm in SQL takes as input a table $G$ with two columns, $v$ and $w$, containing vertex IDs, where each row represents an undirected edge of the input graph. Isolated vertices may be represented in this table as loop edges. The output is a table named \textit{Result} with columns $v$ and~$r$, containing for each vertex $v$ a row assigning a label $r$ to the connected component of~$v$.

Figure~\ref{F:sqlrc} shows an SQL-like pseudocode implementation of Randomised Contraction using the finite fields method. It assumes the existence of a user-defined function $\text{axb}(A,x,B)$ that treats a vertex ID $x$ as an element of a finite field and computes the expression $A\cdot x + B$ using finite field arithmetic. Its implementation along with the actual Python/SQL code used for our experiments is given in Appendix~\ref{A:implementation}.

At each step, the choice of representatives is computed as a table~$R$. For performance optimisation, we compute the representative as $r_i(v) = \min_{w\in N_{G_{i-1}}[v]} h_i(w)$ instead of using $\argmin$. This runs faster because $\min$ is a built-in aggregate function in SQL. Since the values of $r_i$ are no longer vertex IDs of the original graph, the vertices effectively get relabelled at each contraction step. Relabelling does not affect the correctness of the algorithm since the ultimate connected component labels are not required to be vertex IDs, but merely to satisfy uniqueness. Uniqueness is guaranteed by the fact that the functions $h_i$ are bijections on the finite field used as the domain of the vertex IDs.

The contraction step replaces the vertex IDs in each row of the edge table $E$ by their respective representatives, writing the result to a temporary table $T$. This is implemented by joining the edge table $E$ with one copy of~$R$ for each of the two vertices involved. Loop edges are removed from the result to exclude isolated vertices from further computation. 

Recall from section~\ref{S:basic} that the output of the algorithm is the composition of the representative functions $r_k \circ r_{k-1} \circ \cdots \circ r_1$. At step $i$, the algorithm uses the partial composition $r_{i-1} \circ \cdots \circ r_1$ stored in a table $L$ to compute the next partial composition $r_i \circ \cdots \circ r_1$ by joining table $L$ with table $R$. Since isolated vertices get deleted during the course of the algorithm, $R$ represents only a partial function and a left outer join of $L$ and~$R$ has to be used to preserve a row for each of the original vertices. Note that the relabelling introduced by the performance optimisation mentioned above has to be applied to all rows of $L$ that do not have a counterpart in~$R$. This is accomplished using the SQL function coalesce() which returns its first non-NULL argument.

The algorithm in Figure~\ref{F:sqlrc} has deterministic space usage. Table $E$ gets smaller at each step since duplicate edges and loop edges are removed. Table $R$, containing one row per vertex in~$E$, shrinks accordingly. Table $L$, however, maintains its size throughout, storing one row per vertex of the input graph.

\begin{figure}
\begin{algorithmic}
\footnotesize 
\Procedure{RandomisedContractionFast}{$G$}
	\State \textbf{create table E as}
	\State \hskip\algorithmicindent \textbf{select v, w from $G$ union all select w, v from $G$;}
	\State initialise $S$ with an empty stack
	\State $i \gets 0$
	\Repeat
		\State $i \gets i+1$
		\State choose $0 \ne A \in \F$ and $B \in \F$ uniformly at random
		\State push $(A,B)$ onto stack $S$
		\State \textbf{create table $R_i$ as}\Comment{compute representatives}
		\State \hskip\algorithmicindent \textbf{select v, least(axb($A$, v, $B$), min(axb($A$, w, $B$))) as r}
		\State \hskip\algorithmicindent \textbf{from E group by v;}
		\State \textbf{create table T as}\Comment{contract by transforming edge table}
		\State \hskip\algorithmicindent \textbf{select distinct V.r as v, W.r as w}
		\State \hskip\algorithmicindent \textbf{from E, $R_i$ as V, $R_i$ as W}
		\State \hskip\algorithmicindent \textbf{where E.v = V.v and E.w = W.v and V.r != W.r;}
		\State \textit{rowcount} $\gets$ number of rows generated by the previous query
		\State \textbf{drop table E; alter table T rename to E;}
	\Until{$\textit{rowcount} = 0$}
	\State $(A,B) \gets (1, 0)$
	\While{$i>1$}
		\State $i \gets i-1$
		\State pop $(\alpha, \beta)$ from stack $S$
		\State $(A,B) \gets (\text{axb}(A,\alpha,0), \text{axb}(A,\beta,B))$
		\State \textbf{create table T as}\Comment{compose representative functions}
		\State \hskip\algorithmicindent \textbf{select L.v as v, coalesce(R.r, axb($A$, L.r, $B$)) as r}
		\State \hskip\algorithmicindent \textbf{from $R_i$ as L left outer join $R_{i+1}$ as R on ($R_i$.r = $R_{i+1}$.v);}
		\State \textbf{drop table $R_i$, $R_{i+1}$; alter table T rename to $R_i$;}
	\EndWhile
	\State \textbf{alter table $R_1$ rename to Result;}
\EndProcedure
\end{algorithmic}
\caption{A faster version of Randomised Contraction with stochastic space usage. axb is assumed to be a user-defined function that computes the term $A\cdot x+B$ using arithmetic over the finite field~$\F$.}
\label{F:sqlrcfast}
\end{figure}

Figure~\ref{F:sqlrcfast} shows a faster version of Randomised Contraction using slightly more intermediate storage. Instead of joining with the full table $L$ at each step, we first compute and store all representative tables $R_i$. Each one is smaller than the previous one since it contains only one row for each vertex remaining in the computation. In a second loop, these tables are then joined ``back to front'' in a left outer join, again taking the necessary relabelling into account.

The result of both algorithms is $r = r_k \circ r_{k-1} \circ \dots \circ r_1$. The algorithm in Figure~\ref{F:sqlrc} computes $(r_k \circ (r_{k-1} \circ \dots \circ (r_2 \circ r_1)))$ whereas the  algorithm in Figure~\ref{F:sqlrcfast} computes the expression $(((r_k \circ r_{k-1}) \circ \dots \circ r_2) \circ r_1)$. Note, however, that while the algorithm in Figure~\ref{F:sqlrc} guarantees linear space requirements deterministically, the algorithm in Figure~\ref{F:sqlrcfast} only guarantees this in expectation, as shown in Section~\ref{S:space}. The latter algorithm runs faster because it joins the representative tables in small-to-large order whereas the former one joins with the full-size representative table~$L$ at each step.

\section{Performance analysis}\label{S:performance}

\subsection{Time complexity}

The critical observation regarding the Randomised Contraction algorithm is that at each iteration the graph shrinks to at most a constant fraction $\gamma$ of its vertices, in expectation, with $\gamma < 1$. Here we will prove $\gamma\le 3/4$ for the random reals method and the finite fields method. A better bound of $2/3$ is proved in Appendix~\ref{A:bounds} for the case of full randomisation, such as with the random reals method. Note that we only need to consider graphs without isolated vertices since all isolated vertices get removed at the end of each step of the algorithm.

\begin{theorem}
Let $G=\langle V, E \rangle$ be a graph without isolated vertices. For each vertex $v$, let $h(v)$ denote either the random real allotted to $v$ by the random reals method or the integer assigned by the finite fields method. Choose representatives $r(v)=\argmin_{w\in N[v]} h(w)$. Then the expected total number of vertices chosen as representatives is at most $3/4 |V|$.
\end{theorem}

\begin{IEEEproof}
Divide the vertices into high and low vertices according to the median $m$ of the distribution of a random $h(v)$: the high vertices $v$ are those with $h(v)\ge m$.

For a vertex $v$ to choose a high vertex as its representative, it must (1) itself be a high vertex, and (2) have only high vertices as neighbours. Given that $v$ is not isolated, let us pick an arbitrary neighbour of it, $w$, and consider a weaker condition than (2): $w$ must be a high vertex. For the random reals method, both conditions occur independently with probability 1/2. For the finite fields method, let $q=|\F|$. The first condition occurs with probability $\lceil q/2 \rceil / q$ and the second condition, given the first, with probability $(\lceil q/2 \rceil - 1) / q$.

Thus, in expectation, no more than $1/4$ of the vertices choose a high vertex as a representative, proving that in total no more than $1/4 |V|$ high vertices will be chosen as representatives. Even if all low vertices are representatives, this still amounts to an expected number of no more than $3/4 |V|$ representatives in total.
\end{IEEEproof}

Let $\gamma_i$ be the actual shrinkage factor at step~$i$ of the Randomised Contraction algorithm. This is a random variable with $\E(\gamma_i)\le\gamma$. By re-randomising the vertex order at each step, all $\gamma_i$ become independent and therefore uncorrelated. This guarantees that the total shrinkage over the first $k$ steps is in expectation
\[
\E(\prod_{i=1}^k \gamma_i)=\prod_{i=1}^k \E(\gamma_i)\le\gamma^k.
\]

We now show that for any given $\epsilon > 0$ the algorithm terminates with probability $1-\epsilon$ after $\BigO(\log|V|)$ steps. Let $R_k$ be the random variable describing the number of remaining vertices after $k$ steps. The probability of the algorithm not terminating after $k$ steps is $\Pr(R_k\ge 1)$. By Markov's inequality we have $\Pr(R_k\ge1) \le \E(R_k) \le \gamma^k|V|$. Now $\gamma^k|V| \le \epsilon \Leftrightarrow k \ge \log_\gamma\epsilon - \log_\gamma|V| = \BigO(\log|V|)$, which is the desired conclusion.

\subsection{Space requirements}
\label{S:space}

The Randomised Contraction algorithm can be implemented in two variants shown in Figures \ref{F:sqlrc} and~\ref{F:sqlrcfast}, both using the finite fields method. Both require $\Th(|E|)$ space for storing the edge table $E$. Note that the size of this edge table decreases at each step of the algorithm.

The first algorithm uses one table $L$ of size $\Th(|V|)$ and another table $R$ starting at the same size and strictly shrinking throughout the algorithm, so that space usage for these tables is bounded deterministically by $\Th(|V|)$. The algorithm shown in Figure~\ref{F:sqlrcfast} stores intermediate tables of expected sizes $|V|$, $\gamma |V|$, $\gamma^2 |V|$, \dots, $\gamma^k |V|$, which sums up to a space usage of $\Th(|V|)$ in expectation.

If the random reals method is used instead, both algorithms require an additional $\Th(|V|)$ for storing a random number for each vertex, which does not change the overall space complexity.

In summary, since $|V|\le|E|$, the space complexity of the first algorithm is $\Th(|E|)$ deterministically while it is $\Th(|E|) + \text{expected} \Th(|V|)$ for the second algorithm.

In practice, if the algorithms are implemented as shown, the edge table is blown up two-fold in the setup stage. Also, at every iteration, a new edge table has to be generated before the old one is deleted, so, in total, the space requirements for storing edge information during the execution of the algorithm are up to four times the size of its original input.

\section{Empirical evaluation}\label{S:empirical}

\begin{table}
  \caption{Connected component algorithms}\smallskip
  \label{Table:algorithms}
  \centering
  \begin{tabular}{lrr}
    \toprule
    Algorithm&Number of steps&Space\\
    \midrule
    Randomised Contraction\footnotemark{} & exp. $\BigO(\log |V|)$ & exp. $\BigO(|E|)$\\
    Hash-to-Min \cite{Rastogi:MRCC} & $\BigO(\log |V|)$ & $\BigO(|V|^2)$\\
    Two-Phase \cite{kiveris2014connected} & $\BigO(\log^2 |V|)$ & $\BigO(|E|)$\\
    Cracker \cite{lulli2017fast} & $\BigO(\log |V|)$ & $\BigO(\frac{|V|\cdot |E|}{\log |V|})$\\
    \bottomrule
  \end{tabular}
\end{table}
\footnotetext{Space usage can be made deterministic using the implementation in Fig.~\ref{F:sqlrc}.}

To evaluate the practical performance of our Randomised Contraction algorithm we used the open source MPP database Apache HAWQ which runs on an Apache Hadoop cluster. Since SQL does not natively support any control structures, we implemented the algorithm shown in Figure~\ref{F:sqlrcfast} as a Python script that connects to the database and does all the ``heavy lifting'' using SQL queries. Finite field arithmetic over 64-bit integers was implemented in~C as a user-defined SQL function.

We compare Randomised Contraction to three other leading algorithms for calculating connected components in a distributed computation setting. Their proven time and space complexities are summarised in Table~\ref{Table:algorithms}. Hash-to-Min and Two-Phase were implemented by their authors in MapReduce~\cite{miner2012mapreduce} whereas Cracker uses Spark~\cite{karau2017high}.

The use of different execution environments and programming paradigms makes a direct comparison of the algorithms difficult. The authors of Hash-to-Min \cite{Rastogi:MRCC} and Two-Phase \cite{kiveris2014connected} did not publish original code, and comparison difficulties are further exacerbated by the fact that they did not document their cluster configuration and that \cite{kiveris2014connected} provides only relative timing results. We therefore had to port these algorithms to a unified execution environment.

We converted the two MapReduce algorithms and the Spark algorithm to SQL using direct, one-to-one translations. For example, in MapReduce, a ``map'' using key-value messages was converted to the creation of a temporary database table distributed by the key, and the subsequent ``reduce'' was implemented as an aggregate function applied on that table. Spark was converted using an equally direct, straightforward command-to-command mapping. This allows a comparison of different algorithms executing in the same relational database.

For Cracker, we were in addition able to run the original Spark code published in~\cite{lulli2017fast} on our cluster. We also implemented our Randomised Contraction algorithm in Spark SQL. This allows a limited comparison between the two execution environments Spark vs.\ MPP database.

\subsection{Datasets}

The datasets used are summarised in Table~\ref{Table:datasets}. An application to a real-world dataset with nontrivial size is the analysis of the transaction graph of the crypto-currency Bitcoin~\cite{nakamoto2008bitcoin}. At its core, Bitcoin is a data structure called blockchain that records all transactions within the system and is continuously growing.

On April 9, 2019 it consisted of 570,870 blocks with a total size of 250~GB, which we imported into our relational database. Transactions can be viewed as a bipartite graph consisting of transactions and outputs which in turn are used as inputs to other transactions. Each output is associated with an address, and it is a basic step for analysing the cash flows in Bitcoin to de-anonymise these addresses if possible. We used a well-known address clustering heuristic for this~\cite{meiklejohn2013fistful}: if a transaction uses inputs with multiple addresses then these addresses are assumed to be controlled by the same entity, namely the one that issued the transaction. To perform this analysis, we created the graph ``Bitcoin addresses'', linking addresses to the transactions using them as inputs. The connected components of this graph contain addresses assumed to be controlled by the same entities. 

We also calculated the connected components of the full Bitcoin transaction graph. This reveals different markets that have not interacted with each other at all within the crypto-currency.

Another important application of our algorithm is the analysis of social networks. We used the ``com-Friendster'' dataset from the Stanford Large Network Dataset Collection~\cite{snapnets}, the largest graph from that archive.

Connected component analysis can be used as an image segmentation technique. We converted a Gigapixel image ($69{,}536 \times 22{,}230$~px) of the Andromeda galaxy \cite{HubbleGigapixelAndromeda} to a graph by generating an edge for every pair of horizontally or vertically adjacent pixels with an 8-bit RGB colour vector distance up to 50. The vertex IDs were chosen at random so that they would not reflect the geometry of the original image.

\begin{table}
  \caption{Datasets}\smallskip
  \label{Table:datasets}
  \centering
  \begin{tabular}{lrrr}
    \toprule
    Dataset & $|V|$ & $|E|$ & components \\
    \midrule
    Andromeda & 1,459~M & 2,287~M & 62,166~k \\
    Bitcoin addresses & 878~M & 830~M & 216,917~k \\
    Bitcoin full & 1,476~M & 2,079~M & 37~k \\
    Candels10 & 83~M & 238~M & 39~k \\
    Candels20 & 166~M & 483~M & 48~k \\
    Candels40 & 332~M & 975~M & 91~k \\
    Candels80 & 663~M & 1,958~M & 224~k \\
    Candels160 & 1,326~M & 3,923~M & 617~k \\
    Friendster & 66~M & 1,806~M & 1 \\
    RMAT & 39~M & 2,079~M & 5~k \\
    Path100M & 100~M & 100~M & 1 \\
    PathUnion10 & 154 M & 154 M & 10 \\
    \bottomrule
  \end{tabular}
\end{table}

The same technique can be applied to three-dimensional images such as medical images from MRI scans, or to video. We used a 4K-UHD video of a flight through the CANDELS Ultra Deep Survey field \cite{NASACANDELS} and converted some frames of it to a graph using pixel 6-connectivity (x, y, and time) and a colour difference threshold of 20, again randomising the vertex IDs. By using an increasing number of frames we generated a series of datasets (Candels10 \dots\ Candels160) with similar properties and of increasing size for evaluating scalability of the algorithms.

For comparison with~\cite{kiveris2014connected}, we generated a large random graph using the R-MAT method~\cite{chakrabarti2004r} with parameters (0.57, 0.19, 0.19, 0.05), which are the parameters used in~\cite{kiveris2014connected}. Vertex IDs were randomised to decouple the graph structure from artefacts of the generation technique. 

Two worst-case graphs complete our test bench. As shown in the theoretical analysis, Randomised Contraction maintains its logarithmic and quasi-linear performance bounds on any input graph. By contrast, all other algorithms examined have known worst-case inputs that exploit their weaknesses. Path100M is a path graph with 100 million sequentially numbered vertices causing prohibitively large space usage in Hash-to-Min and Cracker. PathUnion10 is the worst case for the Two-Phase algorithm, a union of path graphs of different lengths with vertices numbered in  a specific way.

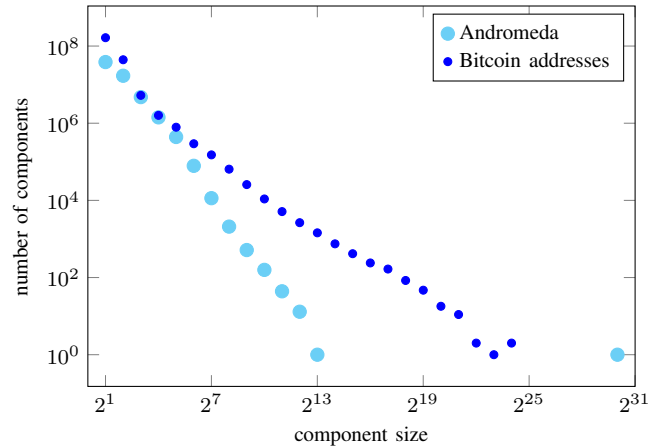
\begin{figure}
  \centering
  \footnotesize
  \begin{tikzpicture}
    \begin{loglogaxis}[
    	height=0.75\columnwidth,
		width=\columnwidth,
    	xlabel={component size},
		ylabel={number of components},
		legend entries={Andromeda,Bitcoin addresses},
		legend cell align={left},
		log base x={2},
		xmin=1,
		xmax=2147483648,
    ]
    \addplot [
    	cyan!50,
		only marks,
    	mark=*,
		mark size=2.5pt,
    ]
	table [col sep=tab]{andromeda-components.dat} ;
    \addplot [
    	blue,
		only marks,
    	mark=*,
		mark size=1.5pt,
    ]
	table [col sep=tab]{address-components.dat} ;
    \end{loglogaxis}
  \end{tikzpicture}
  \caption{Connected component sizes exhibit a roughly scale-free distribution for both the Andromeda and the Bitcoin address datasets.}
  \label{F:ccsizes}
\end{figure}

Our 2D and 3D image connectivity datasets are low-degree graphs: each vertex connects only to a handful of other vertices (at most 4 in 2D, at most 6 in 3D). This is a property that holds in a larger class of graphs of real-world interest, such as, for example, street networks.

With the exception of this degree restriction (for the Andromeda and Candels graphs), however, all graphs in our benchmark exhibit traits that are emblematic of the general class of real-world large graphs, for which reason we are confident that our results are general.

As an example, consider the distribution of our graphs' component sizes. Large real-world graphs typically exhibit a property known as scale-freedom. Scale-freedom in component sizes indicates that on a log-log scale a graph exhibits a (roughly) linear relationship between the size of a component and the number of components of this same size. In Figure~\ref{F:ccsizes}, we demonstrate that the Bitcoin address graph, predictably, shows this log-log linear behaviour.

As can also be seen in Figure~\ref{F:ccsizes}, however, the corresponding plot for the Andromeda benchmark graph shows the same behaviour, so is, in the relevant metrics, also representative of large real-world graphs, despite its construction from an image. (Notably, the single outlier for Andromeda is the image's black background.)

\subsection{In-database benchmark results}

For performance measurements we used a database cluster consisting of five virtual machines, each with 48 GiB of RAM and 12 CPU cores (Intel Skylake @2.2~GHz), running HAWQ version 2.3.0.0 on the Hortonworks Data Platform 2.6.1. The tests were run on an otherwise idle database.

We have run each of the algorithms three times on each of the target data sets and measured the mean and the standard deviation of the computation time. Like any other parallel processing, in-database execution entails its own inherent variabilities, for which reason we did not expect even the deterministic algorithms to complete in precisely consistent run-times. We did, however, expect the randomised algorithm to have somewhat higher variability in its completion time. Observing the relative standard deviation (i.e.\ the ratio between the standard deviation and the mean), the average value for Randomised Contraction was 4.0\% as compared to 2.2\%, 2.1\%, and 1.6\% for Hash-to-Min, Two-Phase, and Cracker, respectively. We conclude that the variability added by randomisation is not, comparatively, very high. 

\begin{table}
  \caption{Runtimes in seconds}\smallskip
  \label{Table:times}
  \centering
  \begin{tabular}{lrrrr}
    \toprule
    Dataset & RC & HM & TP & CR \\
    \midrule
Andromeda & \textbf{5431} & -- & 37987 & 14506 \\
Bitcoin addresses & \textbf{1530} & 11696 & 9811 & 3457 \\
Bitcoin full & \textbf{6398} & -- & 77359 & 26015 \\
Candels10 & \textbf{424} & 3178 & 1425 & 867 \\
Candels20 & \textbf{749} & 5868 & 2836 & 1766 \\
Candels40 & \textbf{1482} & 13892 & 6363 & 3726 \\
Candels80 & \textbf{3463} & -- & 15560 & 8619 \\
Candels160 & \textbf{9260} & -- & 32615 & 23409 \\
Friendster & \textbf{2462} & 9554 & 4409 & 5092 \\
RMAT & \textbf{2151} & 4384 & 2816 & 3187 \\
Path100M & \textbf{366} & -- & 1406 & -- \\
PathUnion10 & \textbf{386} & -- & 4022 & 1202 \\
    \bottomrule
    \noalign{\vskip 3pt}
    \multicolumn{5}{l}{RC = Randomised Contraction, HM = Hash-to-Min}\\
    \multicolumn{5}{l}{TP = Two-Phase, CR = Cracker}

  \end{tabular}
\end{table}

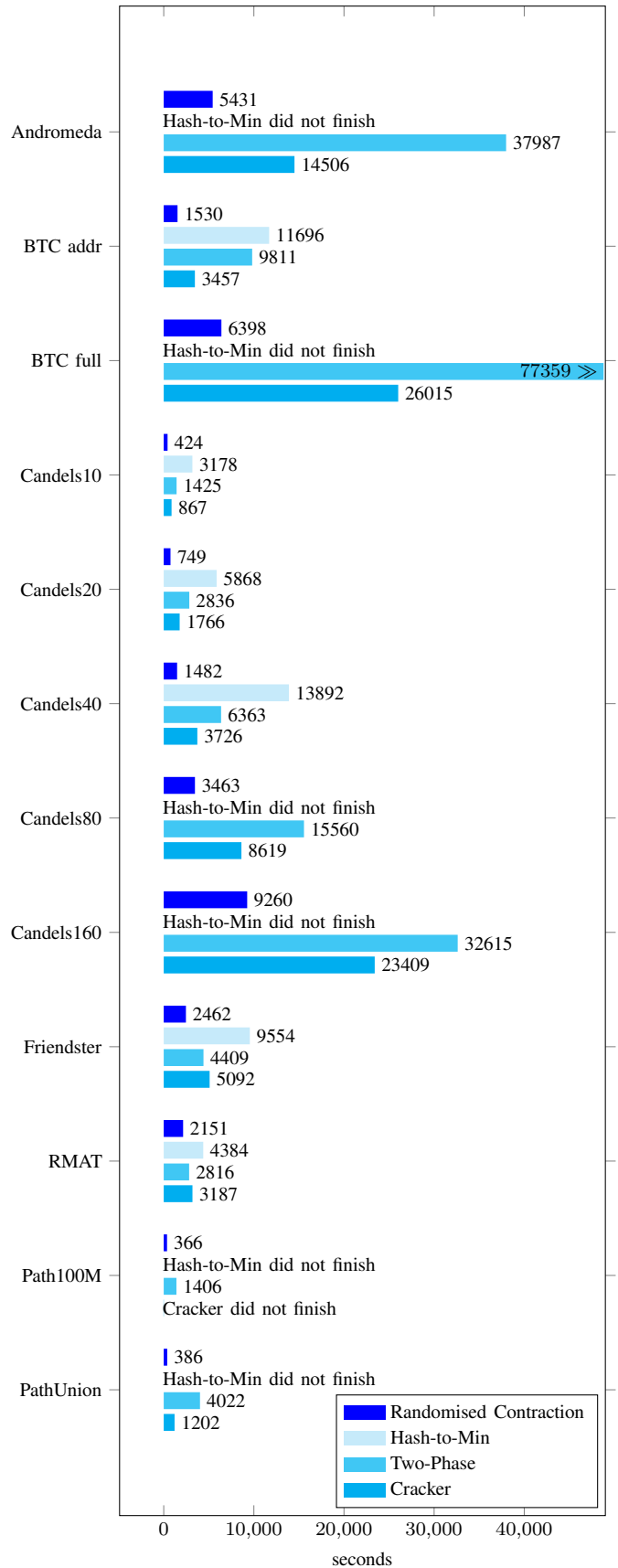
\begin{figure}
  \footnotesize
  \begin{tikzpicture}
    \begin{axis}[
    	width=\columnwidth-6pt,
		height=\textheight-2pt,
		/pgf/bar width=7pt,
    	xbar,
		xlabel=seconds,
		symbolic y coords={Andromeda,Bitcoin addresses,Bitcoin full,Candels10,Candels20,Candels40,Candels80,Candels160,Friendster,RMAT,Path100M,PathUnion10},
        yticklabels={Andromeda,BTC addr,BTC full,Candels10,Candels20,Candels40,Candels80,Candels160,Friendster,RMAT,Path100M,PathUnion},
        y dir=reverse,
		xmax=49000,
		ytick=data,
		nodes near coords,
		nodes near coords align={anchor=west},
		point meta=explicit symbolic,
		scaled ticks=false,
		legend style={at={(0.985,0.005)},anchor=south east},
		legend cell align={left},
		legend image post style={draw opacity=0},
		reverse legend,
    ]
    \addplot [draw=none,fill=cyan,area legend,
    ]
	table [col sep=&,y=dataset,x=CR,meta=xCR]{times.dat} ;
    \addplot [draw=none,fill=cyan!60!white,area legend,
    ]
	table [col sep=&,y=dataset,x=TP,meta=xTP]{times.dat} ;
    \addplot [draw=none,fill=cyan!20!white,area legend,
    ]
	table [col sep=&,y=dataset,x=HM,meta=xHM]{times.dat} ;
    \addplot [draw=none,fill=blue,area legend,
    ]
	table [col sep=&,y=dataset,x=RC,meta=xRC]{times.dat} ;
	
    \legend{Cracker,Two-Phase,Hash-to-Min,Randomised Contraction}
    \end{axis}
  \end{tikzpicture}
  \caption{In-database execution times for real world and synthetic datasets.}
  \label{F:times}
\end{figure}

Table~\ref{Table:times} and Figure~\ref{F:times} show the average runtimes in seconds. Hash-to-Min did not finish on the larger datasets with the available resources. Both Hash-to-Min and Cracker cannot handle the Path100M dataset due to their quadratic space usage (on a shorter path of 100,000 vertices they already use more than 100~GB). On all datasets Randomised Contraction performed best, generally leading by a factor of 2 to~12 compared to the other algorithms. On the graph RMAT the advantage was least pronounced. 

The sequence of Candels datasets, roughly doubling in size from one to the next, demonstrates the scalability of the Randomised Contraction algorithm. Its runtime is essentially linear in the size of the graph.

\begin{table}
  \caption{Maximum space used in GB}\smallskip
  \label{Table:maxspace}
  \centering
  \begin{tabular}{l|r|rrrr}
    \toprule
    Dataset & input & RC & HM & TP & CR \\
    \midrule
Andromeda & 59 & 276 & -- & \textbf{115} & 263 \\
Bitcoin addresses & 21 & 109 & 88 & \textbf{43} & 110 \\
Bitcoin full & 72 & 255 & -- & \textbf{108} & 272 \\
Candels10 & 6 & 27 & 21 & \textbf{12} & 24 \\
Candels20 & 12 & 55 & 42 & \textbf{24} & 50 \\
Candels40 & 25 & 110 & 86 & \textbf{48} & 100 \\
Candels80 & 50 & 221 & -- & \textbf{96} & 201 \\
Candels160 & 102 & 443 & -- & \textbf{193} & 403 \\
Friendster & 47 & 190 & 183 & \textbf{91} & 181 \\
RMAT & 54 & 217 & 120 & \textbf{86} & 169 \\
Path100M & 3 & 13 & -- & \textbf{5} & -- \\
PathUnion10 & 4 & 20 & -- & \textbf{8} & 20 \\
    \bottomrule
  \end{tabular}
\end{table}

\begin{table}
  \caption{Total gigabytes written}\smallskip
  \label{Table:totalspace}
  \centering
  \begin{tabular}{l|r|rrrr}
    \toprule
    Dataset & input & RC & HM & TP & CR \\
    \midrule
Andromeda & 59 & \textbf{552} & -- & 1768 & 905 \\
Bitcoin addresses & 21 & \textbf{215} & 804 & 557 & 306 \\
Bitcoin full & 72 & \textbf{690} & -- & 1858 & 1151 \\
Candels10 & 6 & \textbf{48} & 148 & 93 & 61 \\
Candels20 & 12 & \textbf{97} & 295 & 179 & 125 \\
Candels40 & 25 & \textbf{196} & 618 & 369 & 251 \\
Candels80 & 50 & \textbf{394} & -- & 774 & 504 \\
Candels160 & 102 & \textbf{790} & -- & 1481 & 1009 \\
Friendster & 47 & 309 & 481 & \textbf{258} & 294 \\
RMAT & 54 & 259 & 248 & \textbf{169} & 177 \\
Path100M & 3 & \textbf{31} & -- & 75 & -- \\
PathUnion10 & 4 & \textbf{48} & -- & 264 & 116 \\
    \bottomrule
  \end{tabular}
\end{table}

Real world space usage of the algorithms has two aspects. One is the maximum amount of storage used by the algorithms at any given time, taking into account the amount of space freed by deleting temporary tables. The other, arguably more important metric for database implementations is the total amount of data written to the database while executing the algorithms.

The latter is significant if the whole algorithm is implemented as a \emph{transaction} in a database. A transaction combines a number of operations into one atomic operation that either succeeds as a whole or gets undone completely (\emph{rollback}). In order to achieve this behaviour, most databases delete temporary tables only at the successful completion of the whole algorithm, and therefore storage is needed for all data written during its execution.

Table~\ref{Table:maxspace} shows the algorithms' maximum space usage in comparison with the input size. Here the Two-Phase algorithm uses the least space on all datasets, taking no more than 2 times the storage of the input dataset. Our time-optimised implementation of the Randomised Contraction algorithm stays within the expected bounds and is never more than 2.6 times the space requirements of the Two-Phase algorithm. Table~\ref{Table:totalspace} shows the total amount of data written which would need to be stored in a transaction. Here Randomised Contraction is best in most cases and performs worse only on Friendster and RMAT.

\subsection{Database performance vs. Spark}

In~\cite{lulli2017fast}, Lulli et al.\ implement Cracker, an optimised version called Salty-Cracker, Hash-to-Min, and several other algorithms in the distributed computing framework Spark~\cite{karau2017high}. Their published source code is memory intensive and works within our resources only on smaller graphs. Its execution failed on graphs in our test-bench.

For their most highly optimised version of the Cracker algorithm the dataset with the highest runtime was ``Streets of Italy'' (19~M vertices, 20~M edges). The reported time was 1338 seconds, which was the best among all algorithms compared. We ran our Randomised Contraction algorithm on this same dataset in-database and it finished in 143 seconds. Our database implementation of the Cracker algorithm took 261 seconds.

Note the considerable difference between resources used: the results reported in~\cite{lulli2017fast} were obtained on five nodes with 128~GB of RAM and 32 CPU cores each. Our database cluster had less than half the RAM and half the CPU cores. Also the database was configured as it might be in a real-world production environment, never allocating more than 20\% of the resources to a single query.

Formulating one's algorithm in the form of SQL queries also has advantages beyond in-database execution, as it allows utilising it in other SQL and SQL-like execution environments. As an example, we implemented the Randomised Contraction algorithm in Spark SQL using Spark 2.1.1 and ran it on the Candels10 dataset, exported from the database as a distributed set of text files. This allowed the algorithm to scale up properly, but we note that it was still slower in Spark SQL than when executing in the database. The runtime on our cluster was roughly 2.3 times as long for the Spark SQL implementation as for the in-database one, despite both executing the same SQL code on the same hardware. We conjecture that the main reason for this is the higher level of maturity of the query optimisation that databases such as HAWQ provide.

We note that even this factor of 2.3 does not take into account the amount of time required to export the data from the database for analysis or to re-import the results back into the database, operations that would likely be required in a real world implementation.

\section{Conclusions}

We describe a novel algorithm for calculating the connected components of a graph that can be implemented in SQL and efficiently executed in a massively parallel relational database. Its robustness against worst case inputs and its scalability make it practical for Big Data analytics. The performance measured is not only due to our algorithm's ability to use a minimum number of SQL queries and to minimise the
amount of data handled by each query, but also due to the work of the
database's native, generic query execution optimiser.

With relational databases poised to remain
the standard for storing transactional business data and with query execution
engines improving year to year, the Randomised Contraction algorithm demonstrates
that in-database processing can be a viable and competitive addition to
the more widely used Big Data processing technologies. 

\appendices

\section{Implementation in Python/SQL}
\label{A:implementation}

\begin{figure}
\scriptsize
\begin{lstlisting}[language=C]
/* axplusb(a, x, b) calculates a*x+b over GF(2^64).
   Irreducible polynomial: x^64 + x^4 + x^3 + x + 1
*/
#define IRRPOLY 0x1b

PG_FUNCTION_INFO_V1(axplusb);
Datum
axplusb(PG_FUNCTION_ARGS)
{
  int64 a = PG_GETARG_INT64(0);
  int64 x = PG_GETARG_INT64(1);
  int64 b = PG_GETARG_INT64(2);

  int64 r = 0;
  while (x)
    {
      if (x & 1)
	r ^= a;
      x = (x>>1) & 0x7fffffffffffffff;
      if (a & (1LL << 63))
	a = (a<<1) ^ IRRPOLY;
      else
	a <<= 1;
    }
  PG_RETURN_INT64(r ^ b);
}


\end{lstlisting}
\caption{The user-defined function axplusb.}
\label{F:axplusb}
\end{figure}

\begin{figure}
\scriptsize
\begin{lstlisting}[language=Python]
r.log_exec("setup", 0, """\
create table ccgraph as 
    select v1, v2 from {0}
    union all
    select v2, v1 from {0}
    distributed by (v1);
""".format(dataset))

roundno = 0
stackA = []
stackB = []
while True:
    roundno += 1
    ccreps = "ccreps{}".format(roundno)
    r_A = 0
    while r_A == 0:
        r_A = random.randint(-2**63,2**63-1)
    r_B = random.randint(-2**63,2**63-1)
    stackA.append(r_A)
    stackB.append(r_B)

    r.log_exec("ccreps", roundno, """\
    create table {ccreps} as
        select v1 v,
               least(axplusb({A},v1,{B}),
                     min(axplusb({A},v2,{B}))) rep
        from ccgraph
        group by v1
        distributed by (v);
    """.format(ccreps=ccreps, A=r_A, B=r_B))

    r.log_exec("ccgraph2", roundno, """\
    create table ccgraph2 as 
        select r1.rep as v1, v2
        from ccgraph, {} as r1
        where ccgraph.v1 = r1.v
        distributed by (v2);
    """.format(ccreps))
    r.log_drop("ccgraph")
    
    graphsize = r.log_exec("ccgraph3", roundno, """\
    create table ccgraph3 as 
        select distinct v1, r2.rep as v2
        from ccgraph2, {} as r2
        where ccgraph2.v2 = r2.v
              and v1 != r2.rep
        distributed by (v1);
    """.format(ccreps))
    r.log_drop("ccgraph2")
    r.execute("alter table ccgraph3 rename to ccgraph")

    if graphsize == 0:
        break

accA = 1
accB = 0

while True:
    roundno -= 1
    (accA, accB) = (r.axplusb(accA, stackA.pop(),0), 
                    r.axplusb(accA, stackB.pop(), accB))
    if roundno == 0:
        break
    ccrepsr = "ccreps{}".format(roundno)
    ccrepsr1 = "ccreps{}".format(roundno+1)
    r.log_exec("result", roundno, """\
    create table tmp as
        select r1.v as v, 
            coalesce(r2.rep,axplusb({A},r1.rep,{B})) as rep
        from {r1} as r1 left outer join
             {r2} as r2
             on (r1.rep=r2.v)
        distributed by (v);
    """.format(A=accA, B=accB, r1=ccrepsr, r2=ccrepsr1))
    r.log_drop(ccrepsr)
    r.log_drop(ccrepsr1)
    r.execute("alter table tmp rename to {}".format(ccrepsr))

r.execute("alter table ccreps1 rename to ccresult")
r.log_drop("ccgraph")
\end{lstlisting}
\caption{Our implementation of Randomised Contraction in Python/SQL.}
\label{F:ccrandom}
\end{figure}

In this Appendix we show the implementation of the Randomised Contraction algorithm we used to run the experiments. The user-defined function implementing finite field arithmetic on 64-bit integers in~C is shown in Figure~\ref{F:axplusb}. It is called from SQL as \lstinline$axplusb(A,x,B)$ and computes the expression $A\cdot x+B$.

Our Python code is given in Figure~\ref{F:ccrandom}. It has been stripped of the surrounding infrastructure code. In the excerpt shown, \lstinline$dataset$ contains the name of the input table which is assumed to contain the edge list of the graph in two columns \lstinline$v1$ and \lstinline$v2$, each containing a 64-bit vertex ID.

\lstinline$r.log_exec()$ executes the SQL query passed as the third parameter and returns the number of rows generated. \lstinline$r.log_drop()$ drops the indicated table. \lstinline$r.execute()$ executes miscellaneous SQL queries. \lstinline$r.axplusb(A,x,B)$ calls the corresponding function in the database for finite field arithmetic.

\section{Bounds on graph contraction}\label{A:bounds}

The Randomised Contraction algorithm requires that at each iteration the number of remaining vertices in the graph drops, in expectation, to at most a constant factor $\gamma$ of the initial number, with $\gamma < 1$. In the body of the paper we prove $\gamma\le 3/4$, requiring only the weaker form of randomisation that is achieved by the finite fields method. In this appendix we take a closer look at graph contraction under full randomisation and prove a better bound of $\gamma\le 2/3$ for this case.

To do this, we generalise the problem to directed graphs. We use the following notation~\cite{bang2008digraphs}: let $G = \langle V, A \rangle$ be a directed graph with $n$ vertices. For a vertex $v\in V$, the set $N^+(v)=\{u \mid vu \in A\}$ is called the \emph{out-neighbourhood} of~$v$ and the set $N^-(v)=\{w \mid wv \in A\}$ is called its \emph{in-neighbourhood}. The sets $N^+[v] = N^+(v)\cup\{v\}$ and $N^-[v] = N^-(v)\cup\{v\}$ are called the \emph{closed} out- and in-neighbourhoods, respectively.

We represent an ordering of the vertices by assigning to each vertex~$v$ a unique label $L(v)\in\{1,\dots,n\}$. The representative of a vertex $v$ under the order induced by the labelling~$L$ is defined as $r_L(v) = \argmin_{w\in N^+[v]} L(v)$.

An undirected graph can be considered as a special case of a directed graph where each undirected edge corresponds to a pair of arcs in both directions. In this case we have $N(v) = N^+(v) = N^-(v)$ for all vertices $v$ and our Randomised Contraction algorithm at each iteration chooses representatives as defined above. The total number of distinct representatives then determines the size of the next iteration's graph and therefore the amount of contraction at each iteration. We note that we do not know of any natural interpretation for the result of running the Randomised Contraction algorithm on a directed graph. Certainly, the output is not a division into connected components. 

Given an ordering of the vertices, a vertex can have one of three types: it can be not the representative of any vertex (\textbf{type~0}), the representative of exactly one vertex (\textbf{type~1}), or the representative of two or more vertices (\textbf{type~2+}).

\begin{lemma}\label{L:injective}
Let $G=\langle V,A \rangle$ be a directed graph with $n$ vertices. Fix a vertex $v\in V$ with $N^+(v) \ne \emptyset$. Then the number of orderings under which $v$ is of type~1 is less than or equal to the number of orderings under which it is of type~0.
\end{lemma}

\begin{IEEEproof}
We prove this by constructing an injective mapping from the labellings that make our fixed vertex $v$ a type~1 vertex to those that make it a type~0 vertex. Consider a labelling $L$ that makes $v$ a type~1 vertex. Then there are two cases: (a) $v$ represents itself and (b) $v$ is the representative of exactly one other vertex. 

In case (a) we have $L(v)=\min_{w\in N^+[v]} L(w)$. Let $u_1=\argmax_{w\in N^+(v)} L(w)$ and let $L'$ be a new labelling obtained from $L$ by exchanging the labels of $v$ and~$u_1$. Under this new labelling, $v$ is of type~0: it no longer represents itself and it also does not represent any other vertex because its label is larger than before. Note that we can uniquely identify $u_1$ in this new labelling as $u_1=\argmin_{w\in N^+(v)} L'(w)$.

In case (b) we have $v=r_L(u_2)$ for some vertex $u_2\in N^-(v)$ and $r_L(v) \ne v$. Let $u_1 = r_L(v)$. Then $L(u_2) > L(v) > L(u_1)$. Let $L'$ be a new labelling obtained from~$L$ by exchanging the labels of $v$ and~$u_2$. Under this new labelling, $u_2$ represents itself and $v$ is of type~0: it is no longer a representative of~$u_2$ and it also has not become a representative for any other vertex because its label is larger than before. Furthermore, $L'(u_2) = L(v) > L(u_1) = L'(u_1)$. Note that we can uniquely identify~$u_2$ in this new labelling as the largest-labelled vertex in the in-neighbourhood of~$v$ that represents itself. To see this, assume by contradiction that there is a $w\in N^-(v)$ with $L'(w)>L'(u_2)$ and $r_{L'}(w)=w$. Then $u_2 \notin N^+(w)$, $v \in N^+(w)$, and $L(w)=L'(w)>L'(u_2)=L(v)$. From this and the fact that $L(w) = L'(w) = \min_{x \in N^+[w]} L'(x) \le \min_{x \in N^+[w]\setminus \{v\}} L'(x) = \min_{x \in N^+[w]\setminus \{v\}} L(x)$ we conclude that $r_L(w)=v$. So under the labelling $L$, $v$ is the representative of two distinct vertices $u_2$ and~$w$, contradicting the assumption that it is of type~1.

To see that the mapping from $L$ to $L'$ is injective, it remains to be shown that from $L'$ we can uniquely determine whether it was obtained from case (a) or case~(b). Let $u_1=\argmin_{w\in N^+(v)} L'(w)$ and $u_2 = \argmax_{w \in N^-(v)\colon w = r_{L'}(w)} L'(w)$. If the latter does not exist, $L'$ must have resulted from case (a). We show that otherwise $L'$ satisfies $L'(u_2)>L'(u_1)$ if and only if it is the result of case (b). We have seen in case (b) that $L'(u_2) > L'(u_1)$. In case (a) we have  $L(u_2) = L'(u_2) = \min_{x\in N^+[u_2]} L'(x) \le \min_{x\in N^+[u_2]\setminus\{v,u_1\}} L'(x) = \min_{x\in N^+[u_2]\setminus\{v,u_1\}} L(x)$ and $L(v)<L(u_1)$. If $L(v) < L(u_2)$, this would imply that $r_L(u_2)=v$, contradicting the assumption that $v$ is of type~1. So $L'(u_1) = L(v) > L(u_2) = L'(u_2)$. We conclude that the two cases cannot produce the same labelling and thus our mapping is injective.
\end{IEEEproof}

We can now prove the central theorem of this Appendix.

\begin{theorem}
Let $G=\langle V,A \rangle$ be a directed graph with $n$ vertices and for all $v\in V$, $N^+(v) \ne \emptyset$. Let $L$ be a labelling of $G$ chosen uniformly at random. Then the expected number of vertices chosen as representatives by any vertex satisfies $\E(|\{r_L(v) \mid v\in V\}|) \le (2/3)n$. This is a tight bound.
\end{theorem}

\begin{IEEEproof}
Let $R_0$, $R_1$, and $R_{2+}$ be the expected number of vertices of type 0, 1, and~2+, respectively. From Lemma~\ref{L:injective} we know that for any fixed vertex $v$ its probability of being of type~1 is less than or equal to its probability of being of type~0, since these probabilities are proportional to the corresponding numbers of orderings. This shows $R_1 \le R_0$. Using $R_0 + R_1 + R_{2+} = n$, we get
\begin{equation*}
2R_1 + R_{2+} \le n.
\end{equation*}
By counting the
number of vertices being represented by each vertex we have
\begin{equation*}
R_1+2R_{2+}\le n.
\end{equation*}
Summing the last two equations and dividing by 3 we get
\begin{equation*}
R_1+R_{2+} \le \frac{2}{3} n,
\end{equation*}
which is the desired conclusion
because $R_1+R_{2+}$ is the expected number of representatives.

To prove that the bound is tight, consider that $\gamma=2/3$ is attained
when $G$ is the directed 3-cycle.
\end{IEEEproof}

\begin{figure}
\centering
\begin{tikzpicture}[scale=0.7]
\foreach \x in {0,72,...,288} {
\coordinate (P1) at (\x:1);
\coordinate (P2) at (72+\x:1);
\coordinate (O1) at (57+\x:2);
\coordinate (O2) at (72+\x:2);
\coordinate (O3) at (87+\x:2);
\draw[thick] (P1) -- (P2);
\draw[thick] (O1) -- (P2) -- (O3);
\draw[thick] (P2) -- (O2);
\fill (P2) circle (.1);
\fill (O1) circle (.1);
\fill (O2) circle (.1);
\fill (O3) circle (.1);
}
\end{tikzpicture}
\caption{Graph with highest known contraction factor $\gamma$}
\label{F:max_gamma}
\end{figure}
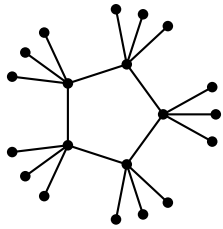

Note that the proven bound is only tight for directed graphs. The worst-case (highest) value of $\gamma$ for undirected graphs is an open question. The graph with the highest
$\gamma$ value known is the one depicted in Figure~\ref{F:max_gamma}. It
has $\gamma=81215/144144 \approx 56.343 \%$.

\section*{Acknowledgment}

Portions of the work described in this paper were done while the second
author was
employed at Pivotal Software, Inc., and are covered by U.~S. Patents.

\bibliographystyle{IEEEtran}
\bibliography{IEEEabrv,bibconnected}

\end{document}